\documentstyle[prc,aps]{revtex}
 
\input epsf 
 
\begin{document}
\draft 

\title{Investigation of low spin states in $^{48}$Cr 
with the MINIBALL spectrometer}

\author{K.~Jessen$^a$, A.~Dewald$^a$, J. Eberth$^a$, G. Gersch$^a$, 
J. Jolie$^a$, N. Pietralla$^a$, T. Steinhardt$^a$, N.~Warr$^a$, 
D. Wei{\ss}haar$^a$, V. Werner$^a$, P.~von Brentano$^a$, 
and S.M. Lenzi$^{b}$}

\address{
$^a$ Institut f\"ur Kernphysik, Universit\"at zu K\"oln, 
Z\"ulpicher Stra{\ss}e 77, D--50937 K\"oln, Germany\\
$^b$ Dipartimentio di Fisica, University of Padova, 
Via F. Marzolo 8, I--35131 Padova, Italy
}

\date{\today}
\maketitle

\begin{abstract}
Low spin states in the self-conjugate even-even nucleus $^{48}$Cr were 
investigated using the {\sc Mini\-ball} $\gamma$-ray spectrometer. At 
the {\sc FN tandem} accelerator in Cologne the 
\mbox{$^{46}$Ti($^{3}$He,n)} reaction was used for the measurement of 
$\gamma\gamma$ coincidences for an excitation function from 7 to 12 MeV 
beam energy. 17 excited states were observed, nine for the first time 
by means of $\gamma$-ray spectroscopy, and new spin assignments were 
made. No excited states apart from the ground band 
were observed below 3.4 MeV.
\end{abstract}

Much experimental and theoretical work has recently been 
devoted to the investigation of self-conjugate nuclei, which 
exhibit unique and interesting properties. For example, they are 
the only nuclei in which isospin zero states ($T=0$), nearly degenerate 
isospin doublets, and large isovector $M1$ transitions can be found. 
In particular the $N=Z$ nucleus $^{48}$Cr, placed amidst the doubly 
magic nuclei $^{40}$Ca and $^{56}$Ni, has been the subject of recent 
studies \cite{cameron,lenzi,brandolini,lis,caurier94,caurier95,sakuda}. 
The ground state rotational band of $^{48}$Cr has been successfully 
described both by the shell model \cite{caurier94} and the collective 
model \cite{caurier95}. The latter approach shows
$^{48}$Cr as a deformed rotor. 
On the other hand the shell model reproduces the states of the 
yrast structure impressively well and predicts many additional 
states above the pairing gap of about 3 MeV. So far these 
additional states had not been observed experimentally. 
The question of their existence puts the shell model to a crucial 
test. A recent $^{40}$Ca$+\alpha$+$\alpha$ cluster model approach 
\cite{sakuda} also describes level energies and $B(E2)$ values of 
the ground band very well and also predicts many so far unobserved 
states.

Early particle-spectroscopic works on the energy levels of $^{48}$Cr
following the (p,t) reaction had shown that
about nine excited states are to be found at level energies 
between 3.4 and 6.1 MeV \cite{bruandet,dorenbusch,shepard}.
Recent measurements used heavy ion induced reactions to 
examine both yrast and non-yrast structures of $^{48}$Cr 
\cite{cameron,lenzi,brandolini}. 
The non-yrast states were assumed to belong to a $K^\pi = (4^-)$ 
band, but the spin and parity assignments had only been tentative 
\cite{brandolini}.
The aim of the present work was to extend the known level 
scheme and remove the spin ambiguity for the side band. 
Therefore we performed a complete $\gamma$ spectroscopy
experiment using the new {\sc Mini\-ball} spectrometer 
\cite{eberth}, which has been designed as an extremely 
efficient array for the detection of low multiplicity 
$\gamma$ events.

The population of non-yrast states in $^{48}$Cr was carried out
following the \mbox{$^{46}$Ti($^{3}$He,n)} reaction at the 
{\sc FN tandem} accelerator of the University of Cologne. 
The {\sc Mini\-ball} spectrometer was used to measure 
$\gamma\gamma$ coincidence events. It was designed to have a
high full-energy peak efficiency and effective granularity, 
needed for the Doppler correction of $\gamma$-rays emitted by 
fast recoiling nuclei. {\sc Mini\-ball} was equipped with 18 
6-fold segmented encapsulated Ge detectors, clustered in six 
Triple-Cluster cryostats with three detectors each. For this 
experiment we used only the total energy signal, which is read  
out from the inner core contact. Neither the segment 
information of the detectors nor a pulse-shape analysis was 
needed, as no Doppler broadening of the $\gamma$ transitions 
was observed in the spectra. All of the detectors were 
positioned as close as possible to the reaction chamber, i.e. 
at a distance of about 10 cm to the target.
A detailed description 
of the detectors and the spectrometer frame can be found in 
Ref. \cite{eberth}.

\begin{figure}[h]
  \epsfxsize 10cm
  \epsfbox{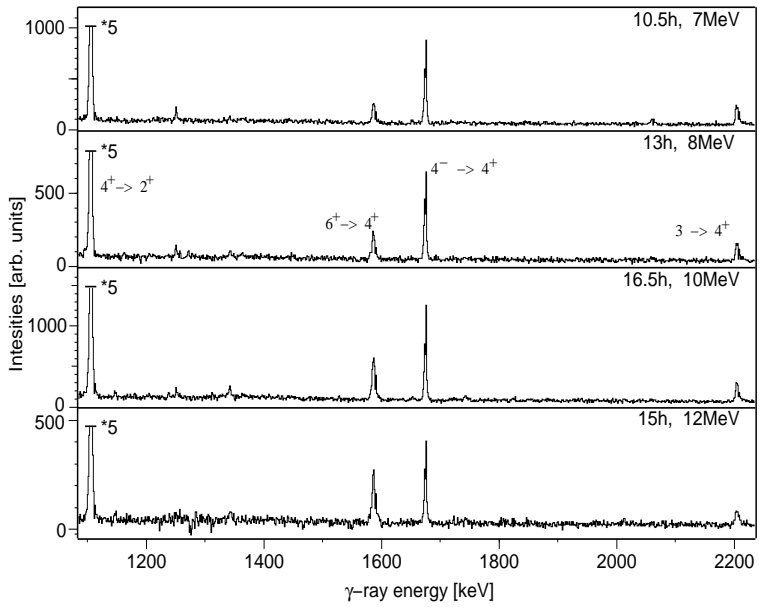}
\caption{Background subtracted coincidence spectra gated by the 
$2^+ \rightarrow 0^+$ transition. The intensity of the $6^+ \rightarrow 4^+$ 
decay increases with the beam energy, contrary to the constancy of the 
$4^- \rightarrow 4^+$ transition. Measuring time and beam energy are 
given in the upper right corner of each spectrum.} 
\label{fig:spc}
\end{figure}

The target consisted of a self-supporting foil of 0.94 mg/cm$^2$ 
$^{46}$Ti , which was bombarded at 7, 8, 10, and 12 MeV beam energy
for a total of about 56 hours measuring time. The sorting of the 
recorded $\gamma\gamma$ coincidences resulted in 
four matrices of 4k$\times$4k resolution, containing 1.9, 2.8, 4.5, 
and $2.9 \times 10^8$ $\gamma\gamma$ coincidence events, 
respectively. The gated spectra in Fig.~\ref{fig:spc} illustrate the 
data quality for each of the different beam energies.

\begin{figure}[h]
  \epsfxsize 11cm
  \epsfbox{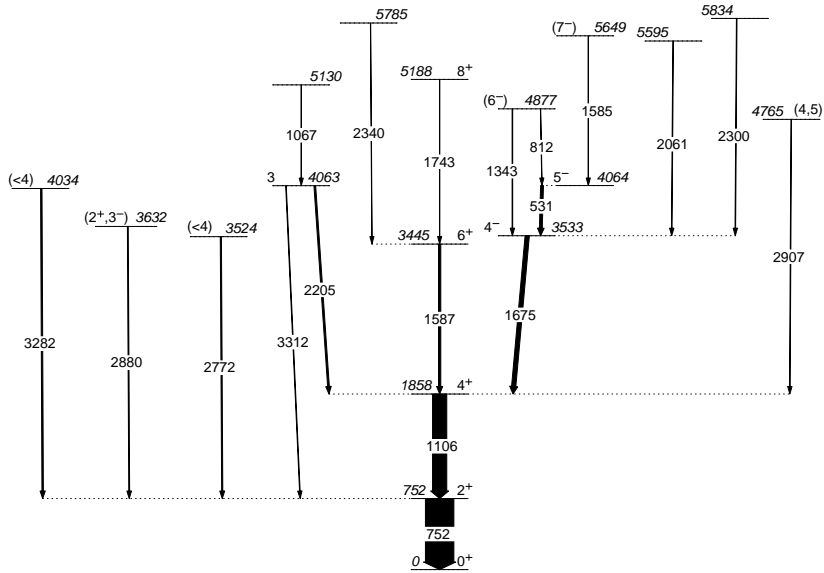}
\caption{Level scheme of $^{48}$Cr from this work. The tentative 
spin value(s) of the 3632 keV level is taken from \protect\cite{NDS} 
and in accordance with our result.}
\label{fig:levelscheme}
\end{figure}

From the $\gamma\gamma$ coincidences the level 
scheme of $^{48}$Cr, shown in 
Fig.~\ref{fig:levelscheme}, was constructed.
With respect to earlier $\gamma$-spectroscopic 
works \cite{cameron,lenzi,brandolini} we observed nine new levels and 
ten new $\gamma$ transitions, exhibiting structures aside from the 
ground state band and 
known non-yrast band. Three levels had already been observed by particle
spectroscopy \cite{dorenbusch}, namely the levels at 3524 and 3632 keV
and the doublet at 4063 and 4064 keV, which had not been recognized as 
such. Our coincidence relations gave clear placements for both levels
of the doublet. No excited states except the $2^+_1$ and $4^+_1$ were 
observed below 3.4 MeV. 
This is a fairly notable result, since the performed $\gamma$-ray 
spectroscopy following the ($^3$He,n) reaction can be regarded as 
complete in the spin range of $3 \hbar$ and $4 \hbar$ up to about
4 MeV excitation energy.

\begin{figure}[h]
  \epsfxsize 8.4cm
  \epsfbox{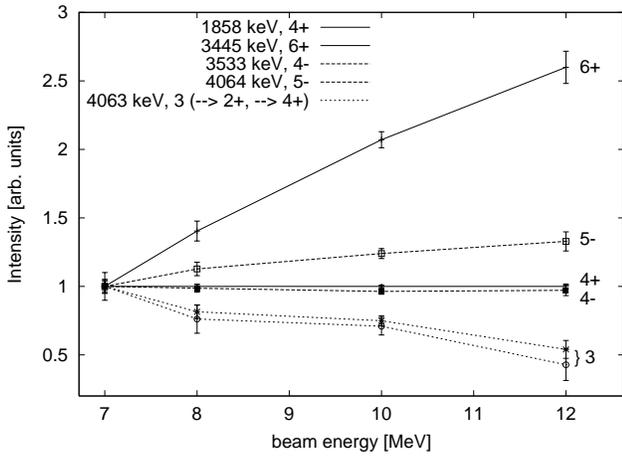}
\caption{Depopulating intensities of different levels (gated from below) 
normalized to the value at 7 MeV and the $4_1^+ \rightarrow 2_1^+$ decay. 
The reference curves for the $4_1^+$ and $6_1^+$ states are drawn solid, 
the curves for the negative parity band are dashed and the ones obtained 
from both decays of the 4063 keV level are dotted.} 
\label{fig:exfunc}
\end{figure}

In order to determine spin values, the excitation function 
for each level was analyzed. We compared the intensities 
of transitions depopulating the levels of interest as functions 
of the beam energy. Those intensities were determined by gating 
from below and normalized to the corresponding intensities at 7 MeV 
beam energy. In a second step the results were normalized to the 
intensities of the decays from the $4^+$ to the $2^+$ state. The 
known spin values of the 1858 and 3445 keV levels, 4 and $6 \hbar$, 
respectively, served as references for the comparison of different 
intensity curves. The results for the states at 3533, 4063, and 
4064 keV are depicted in Fig.~\ref{fig:exfunc}. Our analysis is 
sensitive to 
different spins in the range from $3 \hbar$ to $6 \hbar$. The results 
are summarized in table~\ref{tab:results}. 

Definite spin assignments were made for the states at 3533 and 
4064 keV, which are attributed to the non-yrast band.
The states at 3524, 3632, 4034, and 4063 keV level energy are 
assigned to have a spin of $3\hbar$ or less, but due to the lack 
of known $\gamma\gamma$ excitation functions in this spin range, we 
are not able to give definite assignments.
With the information from an additional 
$\gamma\gamma$ angular correlation experiment, performed with the 
{\sc Osiris-6} cube spectrometer in Cologne, a lower spin limit of
$3 \hbar$ was determined. The experimental details are described in
\cite{JessenPhD}. This measurement gave multipole mixing 
ratios for three levels and branching ratios for the decays of the 
4063 keV level, see table~\ref{tab:results}.

The adopted spin information for the levels that had already 
been observed in \cite{dorenbusch} is in accordance with the 
previous results.
The spin-parity assignment of $J^\pi=3^-$ for one level at 
$(4067\pm 5)$ keV from particle spectroscopy has to be questioned, 
since it is a doublet, that had not been resolved in triton 
spectra.
Nevertheless the spin assignment of $3\hbar$ for the 4063 keV
state is supported by our data. A negative parity assignment 
would be supported by shell model calculations performed with
the code {\sc Antoine} \cite{shellmodelcalc} using the KB3 
interaction, which show no $3^+$ state decaying to the $4_1^+$.

The spin of the non-yrast band head was found to be 
$4\hbar$ and the level directly above this state, connected the 
by 531 keV transition, has a spin of $5\hbar$. 
The multipole mixing ratio for the 1675 keV transition to the 
ground state band vanishes, excluding an appreciable contribution 
of quadrupole radiation. This fact is strongly in favor of 
a negative parity assignment for this $K=4$ band.
Since $^{48}$Cr is a well deformed nucleus
with $\beta \approx 0.3$ for the yrast and non-yrast bands 
\cite{brandolini}, $K$ is regarded as a good quantum number.
In the case of a positive parity band considerable $E2$ character 
and an $E2$ decay to the $2^+_1$ state would have been observed. The 
nanosecond lifetime and pure dipole character of the 
$4^\pi \rightarrow 4^+_1$ decay contradicts the estimated probability 
of a two-fold $K$-forbidden $E2$ transition at least by two orders of 
magnitude. On the other hand, the observations are in good agreement 
with the expected transition probabilities of a three-fold 
$K$-forbidden $E1$, which is isospin forbidden, and a two-fold 
$K$-forbidden $M2$ transition.
Thus, we assign negative parity to this state and the band built 
upon it. Furthermore, no $5^\pi \rightarrow 4^+_1$ was observed.
The upper limit for the ratio of $B(E2)$-values is $B(E2; 5^\pi \rightarrow 4^+_1)/B(E2; 5^\pi \rightarrow 4^\pi) < 2 \cdot 10^{-4}$
and for the intensity ratio
$I_\gamma(5^\pi \rightarrow 4^+_1)/I_\gamma(5^\pi \rightarrow 4^\pi) < 0.1$,
which excludes an appreciable mixing with the ground state band, 
supporting the negative parity assignment to the $K=4$ band. 

\begin{figure}[h]
  \epsfxsize 9.4cm
  \epsfbox{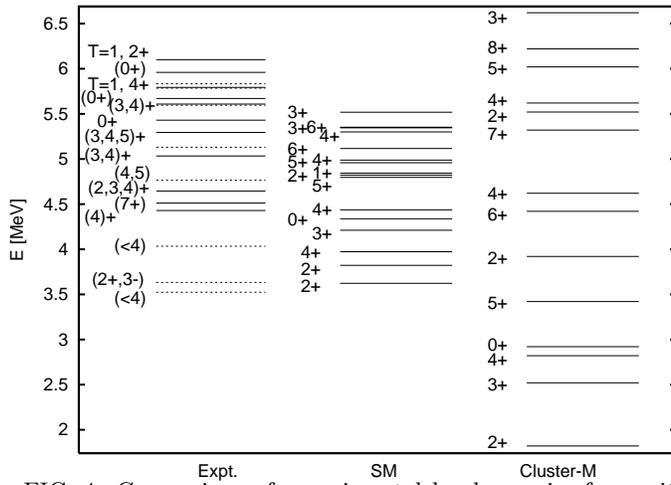}
\caption{Comparison of experimental level energies for positive parity 
states aside from the ground band with the results from our shell model
calculation and from \protect\cite{sakuda}. States of unknown 
parity are drawn with dotted lines.}
\label{fig:EcompPos}
\end{figure}

\begin{figure}[h]
  \epsfxsize 9.4cm
  \epsfbox{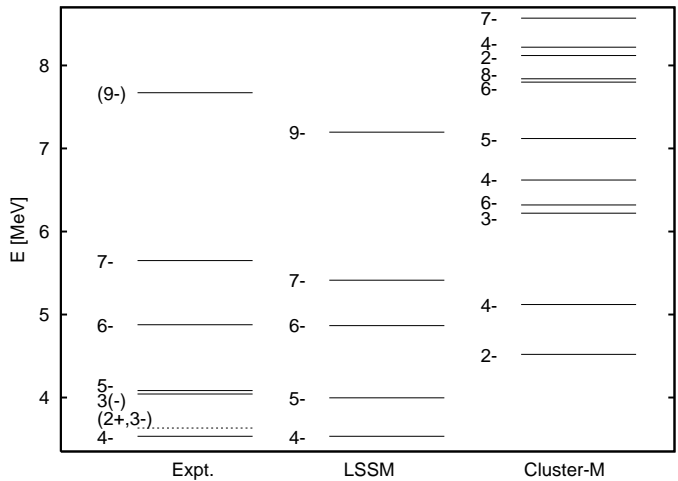}
\caption{Comparison of experimental level energies in the $K^\pi=4^-$ band 
with the results from \protect\cite{brandolini} and \protect\cite{sakuda}.}
\label{fig:EcompNeg}
\end{figure}

In Refs. \cite{brandolini,sakuda} recent shell model and cluster model 
calculations have been found to be in very good agreement with the 
experimental level energies and reduced transition probabilities of 
the $^{48}$Cr ground state band. For positive parity states aside from 
the ground band, our shell model calculations mentioned above
reproduce the energy gap to the first non-yrast state and the 
experimental level density (see Fig. \ref{fig:EcompPos}). A more 
sophisticated comparison is not possible yet, due to the fragmentary 
experimental information on level spins and parities. The calculations 
with the microscopic $^{40}$Ca$+\alpha$+$\alpha$ cluster model by 
Sakuda et al. \cite{sakuda} predict about five positive parity states 
aside from the ground band below 3.4 MeV excitation energy. In the 
present work those states were not observed in the given energy range. 
In particular the $2^+$ band head of a presumed $\gamma$-band must be 
above 3.4 MeV (instead of the calculated 2 MeV).
On the other hand the cluster model predicts the negative parity states 
at rather high excitation energies, e.g. the lowest $4^-$ state above 
6 MeV, which also contradicts the experimental observations. The 
comparison of negative parity level energies is given in 
Fig. \ref{fig:EcompNeg}. In contrast to this the $K^\pi=4^-$ band had 
been well described by Brandolini et al. In their shell model 
calculations the full $pf$-shell had been extended by hole excitations 
in the 1$d_{3/2}$ orbital \cite{brandolini}. 

As a future task 
the collection of more information on the level properties
remains, as the comparison of experimental data with 
the shell model seems promising. Therefore, a further investigation 
of $^{48}$Cr is of high interest, in particular measurements of parities 
for the newly found low spin states.

{\bf Acknowledgment:}
The authors are grateful to the {\sc Mini\-ball} collaboration
for the opportunity to use their powerful instrument. We thank 
A. F. Lisetskiy for fruitful discussions. This work 
was supported by the BMBF project no. 06 OK 958 and the DFG under 
contract numbers Br 799/10-2 and Pi 393/1-2.


\begin{table}[h]
\caption{
New levels, spins, and transitions in $^{48}$Cr from this 
work. The last two columns give the information resulting 
from the additional measurement. See text for details.
}
\label{tab:results}
\begin{center}   
\begin{tabular}{c c c c c} 
$E_x$ & $J$ & $E_{\gamma}$ & $\delta$ & branching \\
 $[$keV$]$ & [$\hbar$] & $[$keV$]$ & & ratio \\
\noalign{\smallskip}\hline\noalign{\smallskip}
 3524 & ($<$4) & 2772 &                         & \\ 
 3533 &      4 & 1675 &              $-0.01(5)$ & \\ 
 3632 & ($<$4) & 2880 &                         & \\ 
 4034 & ($<$4) & 3282 &                         & \\ 
 4063 &      3 & 2205 & $-0.04(5)$ or $\ge 10$  & 100(13) \\
      &        & 3312 &                         & 28(6) \\
 4064 &      5 &  531 &    $0.01(5)$ or $\ge 7$ & \\
 4765 & (4, 5) & 2907 &                         & \\ 
 4877 & (5, 6) & 1343 &                         & \\ 
 5130 &        & 1067 &                         & \\
 5595 &        & 2065 &                         & \\ 
 5785 &        & 2340 &                         & \\ 
 5834 &        & 2300 &                         & \\ 
\end{tabular}
\end{center}
\end{table}

\end{document}